\documentclass[3p,times]{elsarticle}

\usepackage{amsmath, amssymb}
\usepackage{amsthm}

\usepackage{lineno}

\usepackage{color}

\newcommand{\bv}{\mathbf{v}}

\newcommand{\bq}{\mathbf{q}}

\newcommand{\brp}{\bar{p}}

\newcommand{\bx}{\mathbf{x}}
\newcommand{\const}{\mathrm{const}}
\newcommand{\dv}{\mathrm{div\,}}

\newcommand{\pd}[2]{\frac{\partial #1}{\partial #2}}

\newcommand{\prd}[2]{\frac{\partial^2 #1}{\partial #2^2}}
\newcommand\DIV{\text{div}}
\graphicspath{{img/}}

\begin{document}

\begin{frontmatter}




\title{Fluid flow in a reservoir drained by a multiple fractured horizontal well}


\author[address_hydro,address_NSU]{S.V. Golovin\footnote{Corresponding author. E-mail: golovin@hydro.nsc.ru, Address: pr. Lavrentyeva 15, Novosibirsk, 630090, Russia}}
\author[address_NSU]{K.A. Gadylshina}

\address[address_hydro]{Lavrentyrev Institute of Hydrodynamics, Novosibirsk, Russia}
\address[address_NSU]{Novosibirsk State University, Novosibirsk, Russia}

\begin{abstract}
A mathematical model for computation of the fluid pressure in a reservoir drained by a horizontal multiple fractured well is proposed. The model is applicable for an arbitrary network of fractures with different finite conductivities of each segment, for variable in space and time physical parameters of the reservoir and for different field development plans. The variational formulation of the model allows effective numerical simulation using the finite element method. Case studies demonstrate how the main flow characteristics (well productivity, pressure distribution) depend on the geometrical and physical characteristics of the reservoir and of the fracture network. The presented model is suitable for estimation of the productivity of a multiple fractured well and as an optimization tool for efficient reservoir development.
\end{abstract}

\end{frontmatter}


\section{Introduction} Technologies of horizontal drilling and multi-stage fracturing play a key role in a field development for low-permeable reservoirs. Modern advances in engineering allows creation of exact design protocols that consider peculiar properties of the reservoir under development. The appropriate mathematical modelling of future functioning of the multiple fractured horizontal well, estimation of fluid inflow and prediction of the dynamics of the depression zone are important for the rational planning of the field development. The model should take into account geometrical characteristics of the reservoir, variable permeability of the rock, finite hydraulic resistance of fractures and the wellbore, and also be capable to compute the result within a reasonable time on a PC. The prognosis of the inflow and of the geometry of the depression zone can be used as a quick estimation of the quality of the planned wellbore stimulation, or for determining initial data for more advanced industry reservoir simulators.

There are a number of analytical solutions for estimation of productivity of a multifractured horizontal well \cite{2,3,4,5,6}. Although requiring very limited computational resources, these solutions are obtained under restrictive assumptions and simplifications, which limits the area of their application. Hybrid \cite{RenGuo2015,7,8,9,YaoZengLiuZhao2013,Wang2014, XhaoZhangLuoZhang2014} and numerical models \cite{10,11,12,13,14,15} are mostly use not obvious hypotheses regarding the character of the flow near the fractures and the wellbore.  Besides, none of the cited works provide a complete time-dependent pressure field over the reservoir, which might be useful as initial data in commercial simulators.

In present work we develop the hydraulic model proposed in \cite{Kashevarov2010, KuznCherChesnGol2010} by presenting more efficient algorithm of conjugation of flows in the domains of different dimensions and by taking into account variability of physical properties of the reservoir and fractures. The model describes a 3D filtration Darcy flow in the reservoir conjugated with the 2D flow in fractures and with the 1D flow in the wellbore. The conjugation is achieved by a proper choice of conditions over common boundaries of the domains of different dimensions. Our approach is similar to the one presented in \cite{MartJaffRob2005} for description of fractures as interfaces in porous medium. For efficient numerical computations in case of a simplified fractures geometry where all fractures extend from the bottom to the top of the reservoir, we derive a simplified 2D model by averaging pressure along the vertical coordinate. The 2D model has an advantage of faster computation although still reflecting the main geometrical properties of the reservoir.

The model is reformulated in a weak form such that equations for all components of the fluid flow are incorporated into one weak problem suitable for numerical solution by the finite element method. The advantage of the proposed algorithm is that it computes the flow in all segments simultaneously in one time step by the implicit numerically stable scheme. The algorithm also allows taking into account variability of physical properties of the reservoir, finite conductivity of fractures and wellbore. The case studies given in the last section of the paper reveals the dependence of the productivity of the fractured wellbore on the geometrical and physical characteristics of the reservoir and of the fracture network. This analysis demonstrate that the model can be used as an optimization tool for the proper planning of the reservoir development.

\begin{table}
\begin{tabular}{|lp{5.4cm}lp{5.4cm}|}
  \hline
  {\bf Nomenclature} \\
  $p$, $p_0$, $p_w$, $p_\infty$ & pressure, MPa & $\rho$ & density, g/cm$^3$ \\
  $m$, $m_0$ & porosity, $\sim$ & $k$ & permeability, D \\
  $\mu$ & dynamic viscosity, Pa$\cdot$s & $t$ & time, s \\
  $L=L_x$, $L_y$, $H=L_z$ & sizes of the reservoir, m & $\Omega$ & the reservoir \\
   $h_j$, $j=1,\ldots,N$ & fracture heights, cm & $\Sigma$ & outer boundary of $\Omega$ \\

$d_j$, $j=1,\ldots,N$ &fracture aperture, cm& $\Gamma = \bigcup \limits_{i=0}^{N}\Gamma_i$ & inner boundary of $\Omega$ \\
  $k_j$, $j=1,\ldots,N$ & fracture permeability, D & $\theta_j$ & flow rate coefficient, m$^3$s$^{-1}$MPa$^{-1}$\\
  $\gamma_j$, $j=0,\ldots,N$ & imperfection coefficient, $\sim$& $R$ & wellbore radius, m \\
  $\varepsilon$ & the elastic capacity coefficient, MPa$^{-1}$ & $L_w$ & wellbore length, m\\
  $\nu$ & outer normal vector & $\psi$ & test function \\
  $Q_0$ & flow rate at wellbore, g/s & $U$ & pressure within the wellbore, MPa \\
  $\bq_j$, $j=1,\ldots,N$ & total discharge of fluid in fractures, cm$^2$/s& $s$ &local coordinate along wellbore \\
  $q_j$, $j=0,\ldots,N$ & velocity of fluid inflow to the wellbore ($j=0$) and fractures ($j>0$), cm/s &$\alpha_0$ & proportionality coefficient, m$^2$s$^{-1}$MPa$^{-1}$  \\\hline
\end{tabular}
\end{table}

\section {Formulation of the mathematical model}
\begin{figure}
  \centering
  \includegraphics[width=0.6\textwidth]{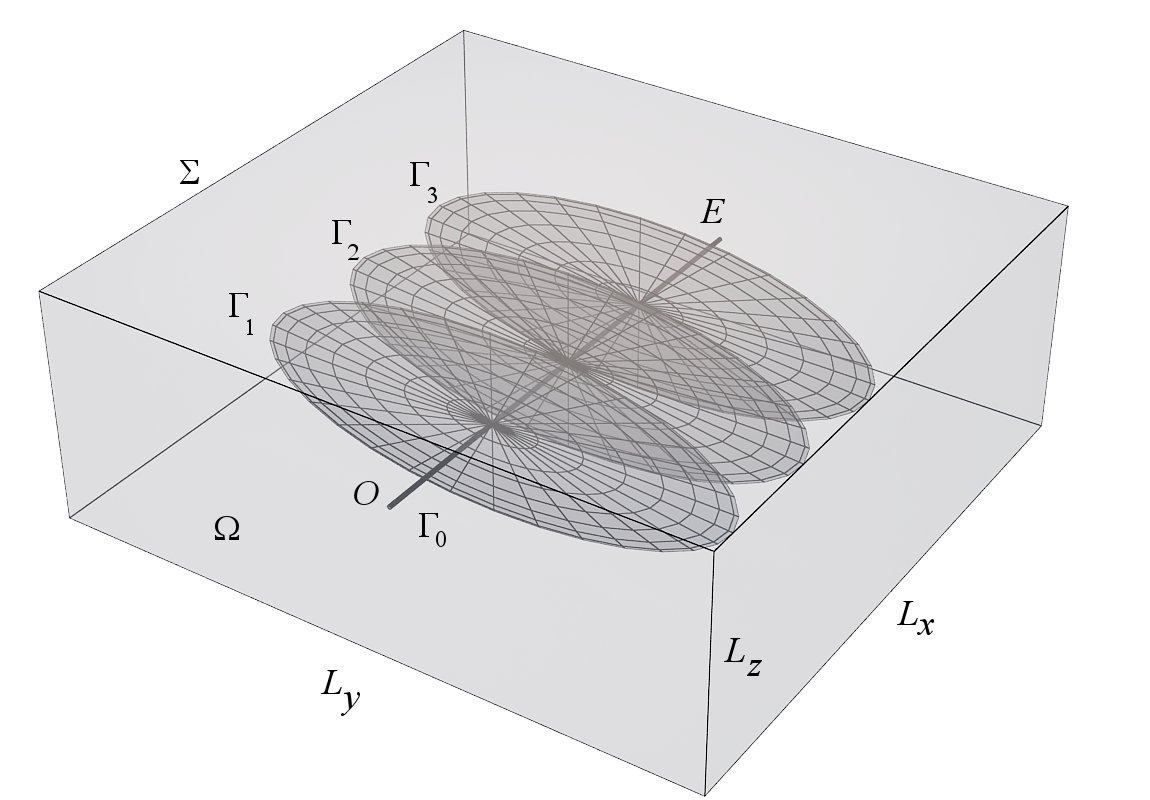}
  \caption{Schematics of the rectangular reservoir $\Omega$ with the outer boundary $\Sigma$ drained by a horizontal wellbore $\Gamma_0$ with multiple fractures $\Gamma_i$ $i=1,\ldots,N$. The borehole is located at point $O$.}\label{WellSketch}
\end{figure}

We observe a single-fluid model of fluid filtration in a rectangular reservoir $\Omega$ with dimensions $L_x \times L_y \times L_z$ (see Figure \ref{WellSketch}). For brevity, we also use the notations $L=L_x$ and $H=L_z$ for the reservoir's horizontal and vertical sizes. The reservoir is exploited by a horizontal well with multiple hydraulic fractures placed arbitrarily along the wellbore. The equation of the one-phase filtration follows from the mass continuity equation and the Darcy law in the form
\begin{equation}\label{eq_1}
\frac{\partial(m\rho)}{\partial t} -\dv \left(\rho \frac{k}{\mu}
\nabla p\right) = 0,\quad \bx\in \Omega \setminus \Gamma.
\end{equation}
Here $\Omega\subset\mathbb{R}^3$ is the modelling domain (the reservoir),  $p$ is the pressure, $\rho$ is the fluid density, $m$ is the porosity, $k$ is the given permeability of the reservoir, $\mu$ is the fluid viscosity. We denote by $\Sigma$ is the outer boundary of $\Omega$, and by $\Gamma = \bigcup \limits_{i=0}^{N}\Gamma_i$ the inner boundaries of $\Omega$, comprising the wellbore $(i = 0)$ and fractures $(1\leqslant i \leqslant N)$. In present work we neglect the gravity force.

In what follows at the modelling of the oil reservoir we neglect the compressibility of the fluid phase $(\rho = \const)$ while taking into account the compressibility of the rock by assuming that the porosity depends linearly on the pressure \cite{Biot1955}:
\begin{equation}\label{poros}
m = m_0\bigl(1 + \varepsilon(p - p_0)\bigr)
\end{equation}
where $m_0$ and $p_0$ are the reference porosity and pressure, and $\varepsilon$ is the elastic capacity coefficient.

It is supposed that the horizontal wellbore $\Gamma_0$ is straight and is parallel to the $x$-axis, has a fixed radius $R$ and given coordinates of the origin $O=(X_w,Y_w,Z_w)$ (the bottom hole) and the end $E$. Hydraulic fractures $\Gamma_j$ $(j=1,\ldots,N)$ are modelled by slots of fixed width $d_j$ as a planar simply connected figures (mostly, rectangles or ellipses), intersecting the horizontal wellbore $\Gamma_0$ at given angles. Position of each fracture $\Gamma_j$ with respect to the wellbore is individual, at that, fractures heights $h_i$ are less or equal to the reservoir thickness $H$.

Interaction of the filtration flow with the well and the fractures is modelled by the boundary condition of the second kind:
\begin{equation}\label{eq_2}
q_j = -\left(\frac{k}{\mu} \frac{\partial p}{\partial \nu}\right)_{\Gamma_j},\quad j=0,\ldots,N.
\end{equation}
Here $q_j$ is the velocity of the inflow to the wellbore or the fracture, $\nu$ is the outer (with respect to the domain $\Omega$) normal to the boundary $\Gamma_j$.

The fluid flow along the wellbore is governed by the continuity equation
\begin{equation}\label{eq_3}
\frac{\partial(S\rho)}{\partial t} + \pd{Q_w}{s} =2\pi R \rho q_0.
\quad 0<s<L_w.
\end{equation}
Here $S=\pi R^2$ is the sectional area of the wellbore, $s$ is the local coordinate along the wellbore, $Q_w=S\rho v_w$ is the flow rate through the section of the wellbore, $v_w$ is the average fluid velocity along the wellbore,  $L_w$ is the length of the wellbore. Assuming laminar flow in the wellbore, the fluid velocity $v_w$ is given by the Hagen-Poiseuille formula
\begin{equation}\label{G-Pw}
v_w=-\gamma_0\frac{R^2}{8\mu}\pd{p}{s}.
\end{equation}
Here $0<\gamma_0 \leqslant 1$ is the imperfection coefficient that takes into account possible additional hydraulic resistance of wellbore due to imperfections. In the general 3-D model we do not distinguish between the reservoir and the wellbore pressure. The model with the individual wellbore pressure that would take into account the wellbore casing or formation of a filter cake on the wall will be observed in a separate paper. Under the assumption of invariability of the wellbore radius and incompressibility of fluid we obtain the relation between the pressure derivatives by the normal and along the wellbore.
\begin{equation}\label{dPw}
\left.\pd{p}{\nu}\right|_{\Gamma_0}=\frac{\gamma_0 R^3}{16 k}\frac{\partial^2 p}{\partial s^2}
\end{equation}

For the flow in hydraulic fractures we use similar conservation laws:
\begin{equation}\label{Cons2D}
\frac{\partial(d_j\rho)}{\partial t} + \dv_2\mathbf{Q_j} =\rho q_j,
\end{equation}
The fluid flow rate is calculated by the formula $\mathbf{Q}_j=\rho \bq_j$ with the total discharge $\bq_j$ given by one of the following relations depending on the problem setup. For clean fracture of width $d_j$ we use the Poiseuille law for the flow in a slot between two plates \cite{Batchelor1967}:
\begin{equation}\label{G-Pw-pl}
\bq_j=-\frac{\gamma_jd_j^3}{12\mu}\nabla_2 p.
\end{equation}
This formula suits well for the injection wells. The case of ``dirty'' fracture filled with proppant is treated by the Darcy law
\begin{equation}\label{Dirt}
\bq_j=-\frac{\gamma_jk_jd_j}{\mu}\nabla_2 p.
\end{equation}
Equation \eqref{Dirt} can be used in case when proppant concentration in the fracture is such that the notion of fracture permeability $k_j$ becomes reasonable. The 2-D differential operators $\dv_2$ and $\nabla_2$ are calculated in local coordinates for each fracture. In both cases $0<\gamma_j\le 1$ is a coefficient characterizing imperfection of fracture walls. Both formulae \eqref{G-Pw-pl} and \eqref{Dirt} are unified by introduction of the flow rate coefficient
\begin{equation}\label{thetadef}
\theta_j=\left\{
\begin{array}{ll}
\frac{\gamma_jd_j^3}{12\mu} & \mbox{ --- clean fracture,}\\[2mm]
\frac{\gamma_jk_jd_j}{\mu} & \mbox{ --- dirty fracture},
\end{array}
\right.\quad j=1,\ldots,N.
\end{equation}

By combining equations \eqref{Cons2D} and \eqref{thetadef} we obtain the expression of the normal (to the fracture's wall) derivative of pressure in terms of derivatives of pressure along the fracture:
\begin{equation}\label{dPf}
\left.\pd{p}{\nu}\right|_{\Gamma_j}=\theta_j\Delta_2 p.
\end{equation}
Here $\Delta_2$ is a 2-D Laplace operator taken in fracture's local coordinates. In the use of formula \eqref{dPf} one should pay attention to the fact  that the fluid inflow should be calculated over both walls of the fracture.

\subsection{Weak formulation of the problem}

Let us write the weak statement of the problem. Due to the incompressibility of fluid $\rho = \const$ and using equation \eqref{poros} we obtain

\begin{equation*}
\frac{\partial m}{\partial
t} = \frac{\partial}{\partial t}\Bigl(m_0(1 + \varepsilon(p - p_0))\Bigr) =m_0 \varepsilon  \frac{\partial p}{\partial t}.
\end{equation*}
Equation \eqref{eq_1} gives

\begin{equation*}
m_0 \varepsilon \frac{\partial p}{\partial t} -\dv\left(\frac{k}{\mu} \nabla p\right)= 0.
\end{equation*}
By multiplication of this equation to an arbitrary function $\psi(\bx)$ and integration over the domain $\Omega$ we find

\begin{multline}\label{varprob}
0=\iiint\limits_{\Omega} \psi  \left(m_0\varepsilon \frac{\partial p}{\partial t}- \DIV\left(\frac{k}{\mu}\nabla p\right)\right)d\Omega =\\
=\iiint\limits_{\Omega} \left(m_0\varepsilon \psi  \frac{\partial p}{\partial t} - \DIV\left(\psi  \frac{k}{\mu}\nabla p\right)
+ \frac{k}{\mu} \nabla\psi  \nabla p\right)d\Omega =\\
= \iiint\limits_{\Omega} \left(m_0\varepsilon \psi  \frac{\partial p}{\partial t} + \frac{k}{\mu} \nabla\psi\nabla p\right) d\Omega -
\frac{k}{\mu} \iint\limits_{\partial\Omega}\psi \frac{\partial p}{\partial \nu}dS
\end{multline}
Let us transform the boundary integral. Over the outer boundary $\Sigma$ depending on the problem's formulation we can state either Neumann or Dirichlet condition, i.e. either the zero flow rate:

\begin{equation*}
\frac{\partial p}{\partial \nu} \Big|_{\Sigma}=0,
\end{equation*}
or a given pressure:

\begin{equation*}
p \Big|_{\Sigma}=p_\infty, \mbox{ at that }
\psi\Big|_{\Sigma}= 0.
\end{equation*}
By using expressions for the normal derivative over the inner boundaries \eqref{dPw}, \eqref{dPf}, we find

\begin{multline}\label{normderiv}
\iint\limits_{\partial\Omega} \psi  \frac{\partial p}{\partial
\nu}dS  =  \iint\limits_{\Gamma_0\setminus\cup(\Gamma_0\cap\Gamma_j)} \psi  \frac{\partial
p}{\partial \nu}dx  + \sum\limits_{j=1}^{N} \iint\limits_{\Gamma_j}
\psi  \frac{\partial p}{\partial\nu}dS =\\
=   \frac{\gamma_0 R^3}{16k} \iint\limits_{\Gamma_0\setminus\cup(\Gamma_0\cap\Gamma_j)}
\psi \frac{\partial^2 p}{\partial x^2}dS   +
\sum\limits_{j=1}^{N}\theta_j \iint\limits_{\Gamma_j}
\psi  \Delta_2 p \,dS =\\
=  \frac{\gamma_0\pi R^4}{8k} \psi
\frac{\partial p}{\partial x} \Big|_{O}^{E}+\frac{\gamma_0\pi R^4}{8k}\sum\limits_{j=1}^{N} \psi
\frac{\partial p}{\partial x}\Big|_{\Gamma_j^+\cap\Gamma_0}^{\Gamma_j^-\cap\Gamma_0}
 - \frac{\gamma_0 R^3}{16k}\iint\limits_{\Gamma_0} \frac{\partial\psi }{\partial
x} \frac{\partial p}{\partial x}dS  +\\
 +  \sum\limits_{j=1}^{N} \theta_j \Big( \int\limits_{\Gamma_j\cap\Gamma_0}\psi
\pd{p}{\nu} dl -  \iint\limits_{\Gamma_j} \nabla_2 \psi\cdot \nabla_2 p\, dS \Big).
\end{multline}
Here by $\Gamma_j^\pm\cap\Gamma_0$ we denote circles obtained at the intersection of the side surface of $j$-th fracture with the cylindrical wellbore $\Gamma_0$. The terms calculated over these intersections, have the meaning of inflows form the fractures to the wellbore, and of the outflows into the wellbore. All together these inflows and outflows compensate each other, hence, the corresponding terms cancel out. At that, integrals over $\Gamma_0$ should be calculated without taking into account the gaps on the intersections of the fractures and the wellbore.

The fluid flow at the end of the fracture $E$ is assumed to vanish, hence,
\begin{equation}\label{Pnorm}
\iint\limits_{\partial\Omega} \psi  \frac{\partial p}{\partial \nu}dS  =  \frac{\mu}{k}
Q_0(t)\psi\Big|_O -\frac{\gamma_0 R^3}{16k} \iint\limits_{\Gamma_0}
\frac{\partial \psi }{\partial
x} \frac{\partial p}{\partial x} dS -  \sum\limits_{j=1}^{N} \theta_j  \iint\limits_{\Gamma_j} \nabla_2 \psi\cdot \nabla_2 p\, dS.
\end{equation}
By integration over $\Gamma_j$ it is necessary to compute integral over both walls of
the fracture.

The variational statement of the problem for the multiple fractured reservoir with the
horizontal wellbore reads as follows:
{\it Find function $p\in W^{1,2}(\Omega)$, satisfying the equation

\begin{multline}\label{weak}
 \iiint\limits_{\Omega} \left(m_0\varepsilon \psi  \frac{\partial p}{\partial t} + \frac{k}{\mu} \nabla\psi\nabla p\right) d\Omega-Q_0(t)\psi\Big|_O +\frac{\gamma_0 R^3}{16\mu} \iint\limits_{\Gamma_0}\frac{\partial \psi }{\partial
x} \frac{\partial p}{\partial x} dS \\
 +  \sum\limits_{j=1}^{N} \theta_j  \iint\limits_{\Gamma_j} \nabla_2 \psi\cdot \nabla_2 p\, dS=0, \quad p|_{t=0}=p_0(\bx)
\end{multline}
for every $\psi(\bx)\in W^{1,2}(\Omega)$ at every time $t\in[0,T]$.}

\subsection{Dimensionless formulation}

It is convenient to introduce the dimensionless variables
\[
t'=t/T, \quad p'=p/p_\ast, \quad q'_0=TQ_0/(LR^2),\quad x'=x/L, \quad y'=y/L,\quad z'=z/L.
\]
Here $T$ and $p_\ast$ are some characteristic scales of time and fluid pressure. Writing the problem \eqref{weak} in dimensionless variables leads to

\begin{multline}\label{weak-dl}
 \iiint\limits_{\Omega'} \left(\psi  \frac{\partial p'}{\partial t'} + a_1 \nabla'\psi\nabla' p'\right) d\Omega' +a_2 \iint\limits_{\Gamma_0'}\frac{\partial \psi }{\partial x'} \frac{\partial p'}{\partial x'} dS'-a_3 q'_0(t)\psi\Big|_{O'} \\
 +  \sum\limits_{j=1}^{N} a_{j+3}  \iint\limits_{\Gamma'_j} \nabla_2' \psi\cdot \nabla_2' p'\, dS'=0, \quad p'|_{t=0}=p'_0(\bx),
\end{multline}
where
\begin{equation}\label{aDl}
a_1 = \frac{Tk}{\mu m_0\varepsilon L^2},\quad
\quad a_2 = \frac{\gamma_0R^3 T}{16\mu L^3 m_0 \varepsilon},\quad
\quad a_3 = \frac{R^2}{L^2 m_0 \varepsilon p_\ast},\quad
a_{j+3}=\frac{\theta_j T}{L^3 m_0\varepsilon}
\end{equation}
In case of non-constant reservoir parameters there will be dimensionless multipliers under integrals in the corresponding terms of equation \eqref{weak-dl}.

The problem under consideration has a small parameter: a ratio $H/L$ of reservoir's width to its length. This observation allows us to construct a simplified 2-D model of the process as described in the subsequent sections.

\section{Two-dimensional model}
For the construction of a simplified 2-D model we will assume that all the hydraulic fractures extend from the bottom to the top of the reservoir: $h_j=H$, $j=1,\ldots,N$. In such a case the reservoir fluid flow is mainly $z$-independent and two-dimensional. Let us introduce the operation of averaging along the vertical axis as
\[\bar{f}(t,x,y)=\frac{1}{H} \int\limits_0^{H} f(t,x,y,z)dz.\]
By substitution of a $z'$-independent test function  $\psi(t',x',y')$ in equation \eqref{weak-dl}, one can perform integration along $Oz'$-axis in integrals over the domain $\Omega'$ as well as in integrals over the fractures $\Gamma'_j$, $j=1,\ldots,N$. As a result of integration, pressure $p'$ transforms to the average pressure $\bar{p}'$ with the multiplier $H/L$. The source term at point $O'$ in the averaged model has the same meaning and does not change. It is remained to average correctly the term, containing the integration over the wellbore $\Gamma_0'$.

In the main volume of the reservoir and in the vicinity of the hydraulic fractures the average pressure $\bar{p}$ does not differ much from the original pressure $p$, but this is not true near the wellbore. At the averaging along the reservoir's hight, the wellbore is substituted by a fictitious fracture extended from the bottom to the top of the reservoir. Parameters of the fictitious fracture should be selected in order to satisfy two conditions:
(a) Equality of inflows of the reservoir's fluid to the fictitious fracture and to the original wellbore; (b) Equality of hydrodynamical resistance for the flow along the fictitious fracture and along the wellbore.

Condition (b) is easily satisfied by a proper choice of the width of the fictitious fracture by equating the flow rates for flow in the fracture and in the wellbore with the same pressure gradient using formulae \eqref{G-Pw} and \eqref{G-Pw-pl}:
\[\frac{\pi R^4}{8\mu}=\frac{Hd_0^3}{12\mu}\quad \Rightarrow\quad d_0=\left(\frac{3\pi R^4}{2H}\right)^{1/3}.\]

On the contrary, condition (a) in case of non-zero inflow can not be satisfied in terms of the present one-pressure model. Indeed, the inflow from the reservoir is determined by the difference of the pressure in the wellbore (fracture) and at some distant point. In case of the wellbore there is a logarithmic singularity at the fracture's axis, whereas in case of the inflow to the fracture the pressure is continuous. This implies, that fluid inflow to the fracture requires much smaller pressure gradient between the fracture and the reservoir than the same inflow to the wellbore. However, due to the coincidence of the hydrodynamical resistance, pressure gradients providing the flow along the fracture and the wellbore are the same. Since the pressure at the origin of the wellbore is known, it is impossible to choose the required pressure along the fictitious fracture that guarantees implementation of both conditions (a) and (b).

For the solution of the declared problem we will consider pressure in the wellbore separately from the reservoir's pressure.

\subsection{Computation of the wellbore pressure}

Let us denote the fluid pressure in the wellbore by symbol $U$. Instead of equation \eqref{eq_2} for $j=0$ we will use the relation
\begin{equation}\label{inflow}
2\pi R q_0 = -2\pi R\left(\frac{k}{\mu} \frac{\partial \brp}{\partial \nu}\right)_{\Gamma_j}=\alpha_0(\brp-U)\Bigr|_{\Gamma_j}
\end{equation}
that declares that the fluid inflow to the wellbore is proportional to the difference between the averaged reservoir's pressure $\bar{p}$ and the wellbore pressure $U$. Coefficient $\alpha_0$ is chosen from the condition of equality of inflows to the planar fracture under the action of the averaged pressure $\bar{p}$, and to the wellbore due to the original 3-D pressure:
 \[\alpha_0=-\frac{2\pi k}{\mu\ln\left(\frac{2R\pi}{H}\sin\bigl(\frac{Z_w\pi}{H}\bigr)\right)}.\]
Details of computation of coefficient  $\alpha_0$ are given in \ref{App1}. Equation \eqref{eq_3} remain unchanged, whereas in equation \eqref{G-Pw} symbol $p$ should be substituted by $U$.

In the new definition of the problem, the fluid flow along the wellbore is driven by the gradient of pressure $U$, which satisfies the continuity equation in the following form
\begin{equation}\label{Ueq}
\pd{}{s}\left(-\frac{\gamma_0\,\pi\,R^4}{8\mu}\pd{U}{s}\right)=\alpha_0\bigl(\brp-U\bigr)+\sum\limits_{j=1}^N\alpha_jd_j(\brp-U)\delta(s-s_j).
\end{equation}
Here by $\alpha_j$ we denote the proportionality coefficients for the wellbore ($j=0$) and fractures ($j=1,\ldots,N$) computed according to   \ref{App1} as follows:
 \[\alpha_j=-\frac{2\pi \theta_j}{d_jp\ln\left(\frac{2R\pi}{H}\sin\bigl(\frac{Z_w\pi}{H}\bigr)\right)},\quad j=1,\ldots,N.\]
Pressure $\brp$ in equation \eqref{Ueq} should be taken at the corresponding point of the wellbore. Values $s_j$ in the argument of the Dirac delta-function $\delta$ are coordinates of the hydraulic fractures in the local coordinates along the wellbore.

Boundary conditions to equation \eqref{Ueq} can be chosen either as a given flow rate or a given pressure at the origin $O$. At the end $E$ of the wellbore we assume zero flow rate:
\begin{equation}\label{Uboundrate}
\left.\pd{U}{s}\right|_{s=0}=-\frac{8\mu Q(t)}{\pi R^4\gamma_0}, \mbox{ or } U|_{s=0}=p_w(t);\quad\left.\pd{U}{s}\right|_{s=L_w}=0.
\end{equation}
Let us choose the following dimensionless variables:
\[U'=U/p_\ast, \quad \delta' = L\delta.\]
Equation \eqref{Ueq} and boundary conditions \eqref{Uboundrate} in the dimensionless variables take the form (primes are omitted):
\begin{equation}\label{Ueqdl}
\begin{array}{l}
\displaystyle -\prd{U }{s }=b_1(\brp -U )+\sum\limits_{j=1}^Nb_{j+1}(\brp-U)\delta(s-s_j),\\[4mm]
-\displaystyle \left.\pd{U }{s }\right|_{s =0}=b_0 q _0(t),\quad U |_{s =L _w}=\brp
\end{array}
\end{equation}
Dimensionless parameters $b_j$ have the following expressions in terms of the dimensional variables:
\[b_0=\frac{8\,\mu\,L^2}{\gamma_0\,\pi\,R^2\,T\,p_\ast},\quad b_1=\frac{8\,\mu\,L^2\,\alpha_0}{\gamma_0\,\pi\,R^4},\quad b_{j+1}=\frac{8\,\mu\, d_j\,L\,\alpha_j}{\gamma_0\,\pi\, R^4}.\]
Equation \eqref{Ueqdl} describes the pressure distribution in the wellbore and takes into account the fluid inflow from the reservoir and from the fractures.

\subsection{Weak form of the 2-D model}

For the numerical calculations it is convenient to use the weak form of the problem. Formula \eqref{Pnorm} for the fluid inflow to the wellbore and to the fractures is modified as follows:
\begin{equation}\label{Pnorm2D}
\iint\limits_{\partial\Omega} \psi  \frac{\partial \brp}{\partial \nu}dS  =  -\frac{\mu}{k}
\int\limits_{O}^E\alpha_0\psi(\bar{p}-U)ds
-\frac{\mu}{k}\sum\limits_{j=1}^N\alpha_jd_j\psi(\bar{p}-U)|_{\Gamma_j\cap\Gamma_0}
-\sum\limits_{j=1}^{N} \theta_j H  \int\limits_{\Gamma_j} \pd{\psi}{s}\pd{\bar{p}}{s}\, ds.
\end{equation}
Thus, we obtain the following equation for the averaged pressure $\bar{p}$ in the dimensionless variables (primes are omitted):
\begin{equation}\label{weak-dl-2D}
 \iint\limits_{\Omega_2 } \left(\psi  \frac{\partial \bar{p} }{\partial t } + a_1 \nabla \psi\cdot\nabla  \bar{p} \right) d\Omega
 +a_2\int\limits_{O}^E\psi(\bar{p}-U)ds
 +  \sum\limits_{j=1}^{N}\left(a_{2j+1}\psi(\bar{p}-U)|_{\Gamma_j\cap\Gamma_0}+ a_{2j+2}  \int\limits_{\Gamma _j} \pd{\psi}{s}\pd{\bar{p}}{s} \, ds\right) =0,
\end{equation}
where
\begin{equation}\label{aDl-2D}
a_1 = \frac{Tk}{\mu m_0\varepsilon L^2},\quad a_2 = \frac{\alpha_0 T}{m_0 \varepsilon H L},\quad
a_{2j+1} = \frac{\alpha_jd_j T}{m_0 \varepsilon HL^2},\quad a_{2j+2}=\frac{\theta_j T}{L^3 m_0\varepsilon}.
\end{equation}
Here by $\Omega_2$ and $\Gamma_j$, $j=0,\ldots,N$ we denote projections of domain $\Omega$, of the wellbore and of the fractures to $Oxy$ plane, represented by a rectangle and lines respectively.
Equation \eqref{weak-dl-2D} can be modified with the help of equation \eqref{Ueqdl} as
\begin{equation}\label{weak-dl-2D-1}
 \iint\limits_{\Omega_2 } \left(\psi  \frac{\partial \bar{p} }{\partial t } + a_1 \nabla \psi\nabla  \bar{p} \right) d\Omega
 +c_1\int\limits_{O}^E\pd{\psi}{s}\pd{U}{s}ds-c_0q(t)\psi(O)
 +  \sum\limits_{j=1}^{N}a_{2j+2}  \int\limits_{\Gamma _j} \pd{\psi}{s}\pd{\bar{p}}{s} \, ds =0,
\end{equation}
where
\begin{equation}\label{aDl-2D-1}
c_0 = \frac{R^2}{m_0\varepsilon H L p_\ast},
\quad c_1 = \frac{\gamma_0\pi R^4 T}{8\mu m_0 \varepsilon H L^3}.
\end{equation}
Multiplication of equation \eqref{Ueqdl} to the test function $\eta$ and integration on $s$ from $O$ to $E$ with the use of the boundary conditions \eqref{Uboundrate} leads to
\begin{equation}\label{weak-U-2D}
\begin{array}{l}
\displaystyle \int\limits_{O}^{E}\pd{\eta}{s}\pd{U}{s}ds-b_1\int\limits_{O}^{E}\eta(\bar{p}-U)ds- \sum\limits_{j=1}^Nb_{j+1}(\bar{p}-U)\eta|_{s=s_j}-b_0 q_0(t) \eta|_{s=0}=0.
\end{array}
\end{equation}

The weak formulation of the 2-D problem is now states as follows: {\it Find function $p\in W^{1,2}(\Omega_2)$ and function $U\in W^{1,2}(\Gamma_0)$ satisfying equations  \eqref{weak-dl-2D} (or \eqref{weak-dl-2D-1}) and \eqref{weak-U-2D} for every functions $\psi\in W^{1,2}(\Omega_2)$ and $\eta\in W^{1,2}(\Gamma_0)$ at every time $t\in[0,T]$}.

\section{Case studies}

The purpose of the following numerical experiments is to demonstrate the agreement of our numerical results with known data and adequacy of the behaviour of the main flow parameters as functions of the geometry and physical properties of the problem.

\subsection{Numerical realization}
Numerical calculations are performed in a freely accessible finite element solver FreeFEM++ \cite{FreeFem}. For the calculations we use weak dimensionless formulation \eqref{weak-dl-2D}--\eqref{weak-U-2D}. Symmetry of the problem with respect to the unknown functions $(p, U)$ and test functions $(\psi, \eta)$ assures the symmetry of the stiffness matrix which is important for the correctness of the numerical algorithm.

Time derivative is approximated by a finite difference: $\partial p/\partial t\approx (p^{n+1}-p^n)/\tau$
where $\tau$ is the time step. The upper index denotes the time instant: $p^n=p(t_n, \mathbf{x})$, $t_n=n\tau$.
Time iterations are performed using the first-order backward (implicit) Euler method. Due to unconditional stability of the method, the value of
the time step $\tau$ does not depend on the characteristic diameter of the mesh cells.

For the construction of a numerical mesh we take 200 mesh vertices on the outer border $\Sigma$, and a proportional number of vertices over the wellbore and fractures. The total number of mesh vertices varies from 6000 to 13000 depending on the complexity of the geometry. Dimensionless time step is equal to 0.05 which is equivalent to 1 hour 12 minutes. In all cases time of calculation on 4000 time steps was about 10--15 of minutes on a usual PC.

\subsection{Common parameters}

The main reservoir and fluid parameters used in numerical tests are listed in Table \ref{ParametersSet}. In all tests we prescribe the borehole pressure $p|_O=p_w$ and calculate the volume of produced fluid using formula \eqref{weak-U-2D} with $\eta=1$. Unless otherwise mentioned, we use the Neumann's (no-flow) outer boundary condition: $\partial p/\partial n|_\Sigma=0$. Initial pressure in undisturbed reservoir is assumed to be zero: $p(0,\bx)=0$. In our model there is no difference between production and injection wells, therefore for the sake of convenience we assume positive pressures by taking $p_w>0$.

\begin{table}[t]
	\caption{Parameters set for the reservoir}
	\label{ParametersSet}
	\begin{center}
	\begin{tabular}{l c l}\hline\hline
 		Parameter & Symbol &  \multicolumn{1}{c}{Value} \\ \hline
        reservoir size & $L_x\times L_y\times L_z$ & $2800  \times 2600 \times 20$ m \\
        wellbore origin & $O=(X_w,Y_w,Z_w)$ & $(800,1300,10)$ m\\
        wellbore radius & $R$ & 10 cm \\
        wellbore length & $L_w$ & 1100 m \\
        fracture aperture & $d_j$ & 1 cm, $j=1,\ldots,N$ \\
        fracture permeability & $k_j$ & 1000 D $\approx0.987\cdot 10^{-9}\mathrm{m}^2$, $j=1,\ldots,N$\\
        borehole pressure & $p_w$ & 10 MPa \\
        pressure at infinity & $p_\infty$ & 0 MPa\\
 		fluid dynamic viscosity & $\mu$ & 1 cP $\approx 10^{-3} \mathrm{Pa}\cdot\mathrm{sec}$\\
 		porosity & $m_0$ & 0.1 \\
 		elastic capacity coefficient & $\varepsilon$ & 1 GPa$^{-1}$ \\
 		imperfection coefficient & $\gamma_j$ & $\in (0, 1]$, $j=0,\ldots,N$ \\
		\hline
	\end{tabular}
	\end{center}
\end{table}

\begin{figure}[t]
    \centering
  \includegraphics[width=0.5\textwidth]{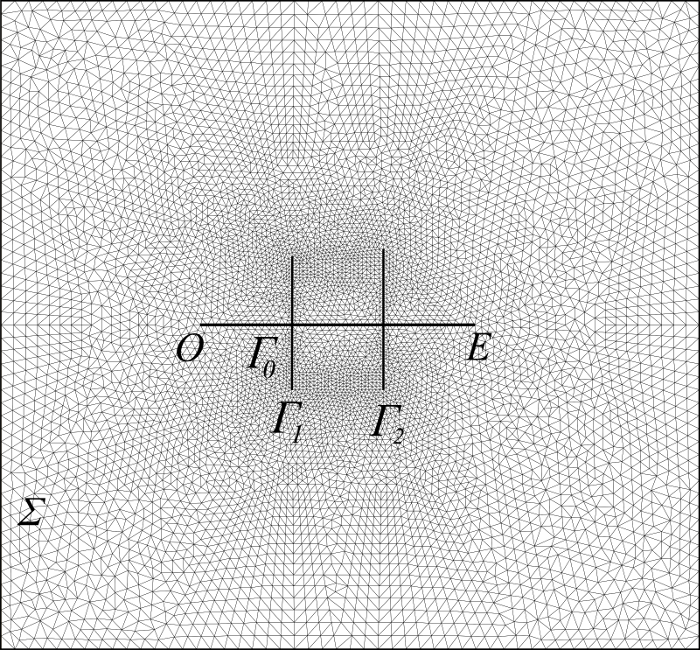}\\
  \caption{Positions of the wellbore and fractures in the reservoir with the computational mesh}\label{Case1fracs}
\end{figure}

\begin{table}[t]
	\caption{Set of instrumental parameters for Case 1}
	\label{DataSet}
	\begin{center}
	\begin{tabular}{l p{4 cm} p{4 cm} p{4 cm}}
		\hline\hline
 		index & rock permeability $k$, mD & imperfection coefficient $\gamma_0$ & number of fractures $N$\\
		\hline
 		0 & 1  & 1 & 0  \\
        1 & 10 &  $3\cdot 10^{-5}$ & 2 \\
		\hline
	\end{tabular}
	\end{center}
\end{table}

\subsection{Case 1: Influence of rock permeability, well conductivity and presence of fractures}

We begin with the test that demonstrate the influence of the rock permeability, wellbore conductivity and presence of fractures to the well production. We number the observed cases by a 3-digits binary number as indexed in Table \ref{DataSet} where the instrument parameters are listed. For example, the case \#101 corresponds to $k=10$ mD, $\gamma_0=1$, $N=4$. In cases of fractured wellbore we consider two linear fractures intersecting the wellbore at right angle as shown in Figure \ref{Case1fracs}. The first fracture has symmetrical wings of length 260 m each, the second fracture has non-symmetrical wings of 260 an 305 m. The imperfection coefficients are $\gamma_1=0.5$ for the first fracture and $\gamma_2=1$ for the second fracture.

\begin{figure}[t]
    \centering
  \includegraphics[width=0.8\textwidth]{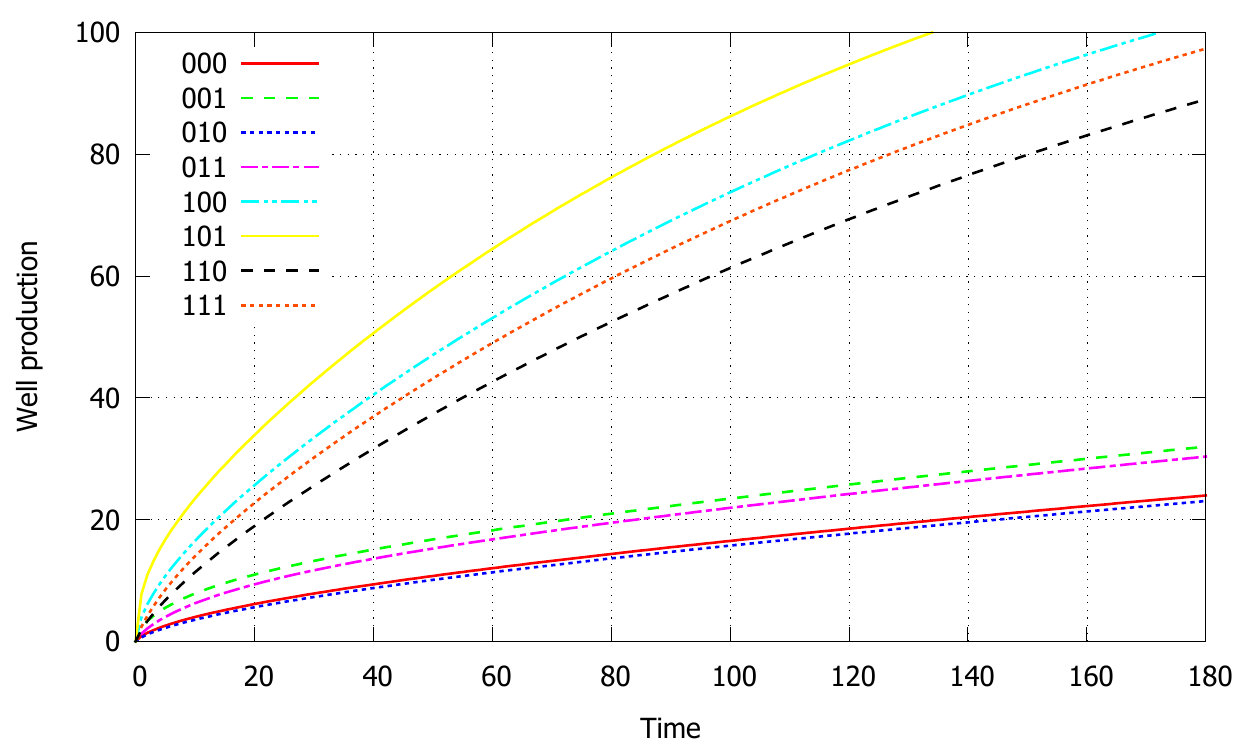}\\
  \caption{Well production ($10^3$ m$^3$) versus time (days) for various geometrical setups. Values of the instrument parameters are coded by a 3-digits binary number according to Table \ref{DataSet}}\label{Case1}
\end{figure}

The calculated volume of produced fluid in all 8 cases during the period of 180 days is shown in Figure \ref{Case1}. Higher productivity is expectably obtained in cases with hydraulic fracture \#xx1 in contrast to cases of a single wellbore \#xx0.

Lower values of imperfection coefficient $\gamma_0$ correspond to higher hydraulic resistance of the wellbore which limits the volume of produced fluid. Difference in the wellbore conductivity affects the well productivity weakly in cases of the lower rock permeability (pairs 000 --- 010, 000 --- 011) and strongly in cases of higher permeability (pairs 100 --- 110, 101 --- 111). The influence of the rock permeability is also expectable: higher volumes of fluid are obtained for the higher permeabilities.

Figure \ref{Case1} contains graphs that are drawn with the same data as some cases observed in \cite{Kashevarov2010}, namely, cases \#110 and \#111 correspond to graphs number 2 in Figures 2 and 3 in \cite{Kashevarov2010} respectively. One can see a good agreement between these pairs of graphs which confirms the reliability of our numerical algorithm.

\subsection{Case 2: Influence of the relative positions of fractures}
In this set of cases we vary the number of fractures, lengthes and distances between fractures. We observe three cases: (a) four symmetrical fractures on the equal intervals of 220 m, each of the total length (both wings) of 1040 m; (b) two fractures separated by the intervals of 367 m, each of the total length 2080 m; (c) four symmetrical fractures separated by the shorter interval of 110 m of the total length 1040 m each. In all cases the rock permeability is $k=1$ mD and imperfection coefficients are $\gamma_j=1$, $j=0,\ldots,4$, The idea is to compare the productivity of fractures of the same total length 4160 m in different geometrical setups.

\begin{figure}[t]
    \centering
  \includegraphics[width=0.8\textwidth]{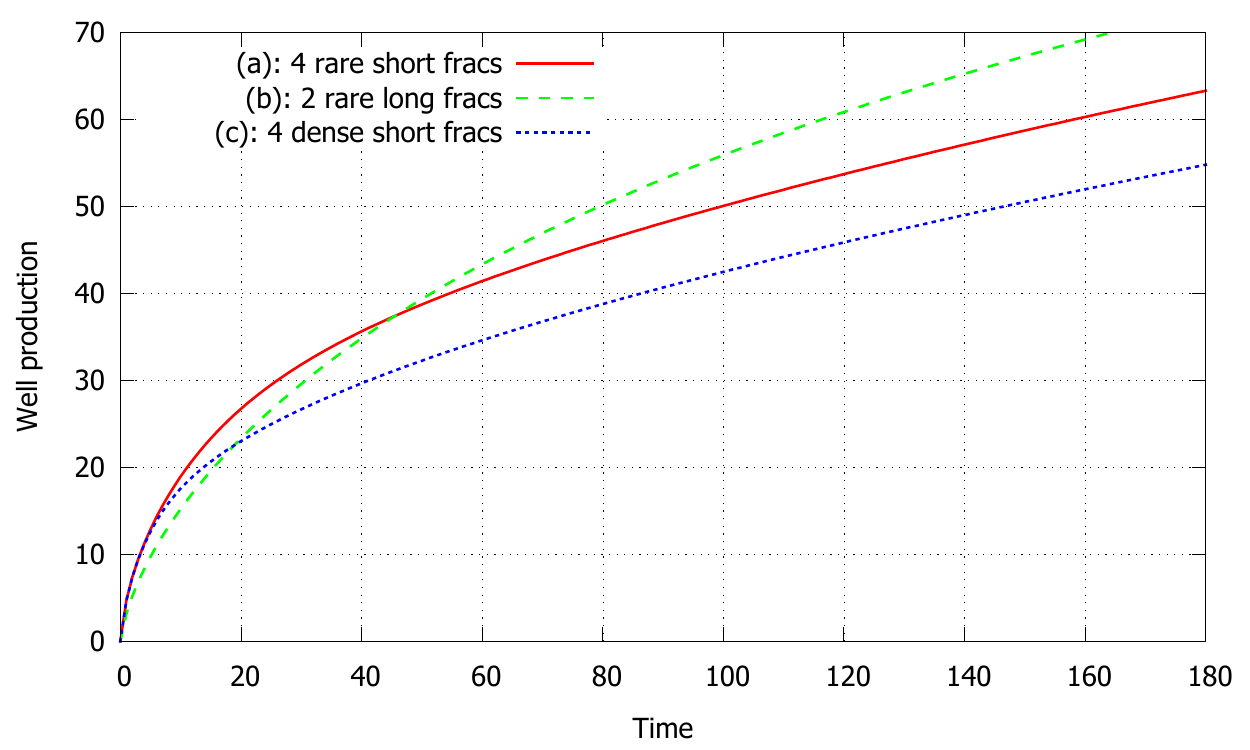}\\
  \caption{Well production ($10^3$ m$^3$) versus time (days) for different geometrical setups: (a) --- four short rare fractures; (b) --- two long fractures; (c) --- four dense short fractures}\label{Case2}
\end{figure}

\begin{figure}[t]
    \centering
  \includegraphics[width=0.3\textwidth]{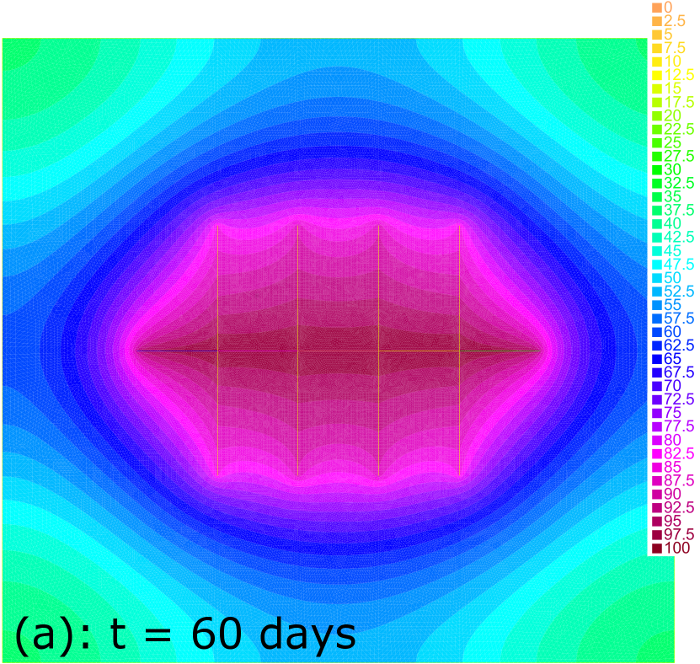}\hfill
  \includegraphics[width=0.3\textwidth]{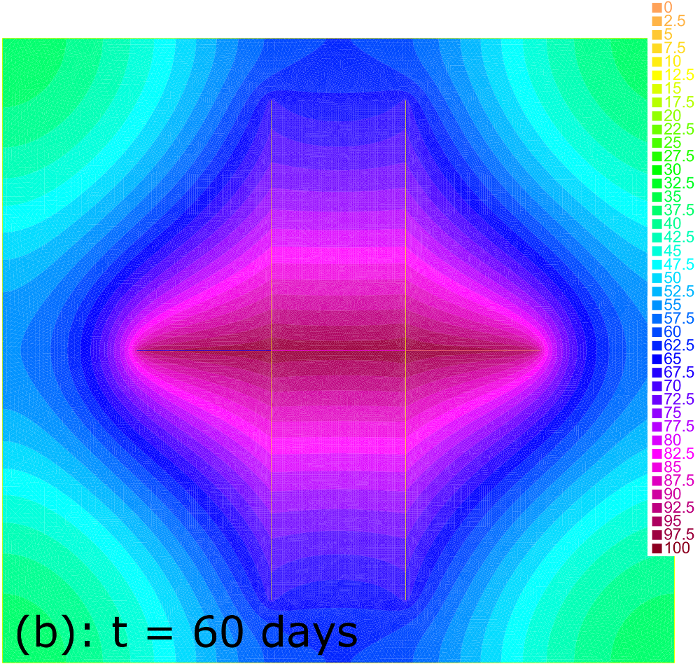}\hfill
  \includegraphics[width=0.3\textwidth]{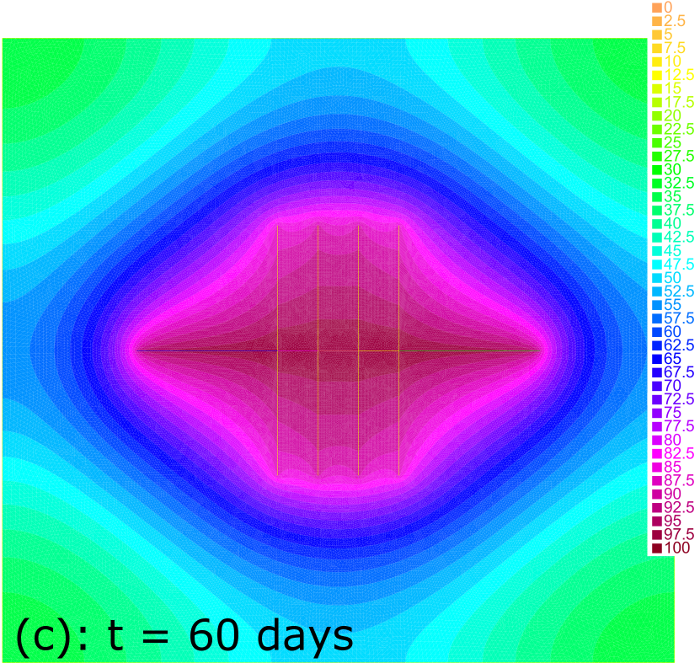}
  \caption{Pressure distribution for different geometries of fractures at $t=60$ days.}\label{Case2press}
\end{figure}

Figure \ref{Case2} demonstrates the volume of produced fluid versus time in each of the observed cases. One can see that cases (a) and (c) do not distinguish within first 10 days of production until the drainage zones of fractures start to influence each other. In case (c) of denser fractures the mutual influence of the drainage zones decreases the productivity in comparison to case (a) of fractures placed more rare. In contrast, two long fractures in case (b) of the same total length as in case (a) are less productive during first 40 days due to the finite conductivity of fractures. However, for longer time two long fractures are more productive than 4 shorter ones due to the larger area of the drainage zone. Pressure distribution for all three cases at $t = 60$ days is shown in Figure \ref{Case2press} where the difference of the drainage zones is clearly seen.
Thus, this example demonstrate the potential of the model for optimization of the fractures network for given reservoir conditions.

\subsection{Case 3: An ``arbitrary'' fracture net}

\begin{figure}[t]
    \centering
  \includegraphics[width=0.6\textwidth]{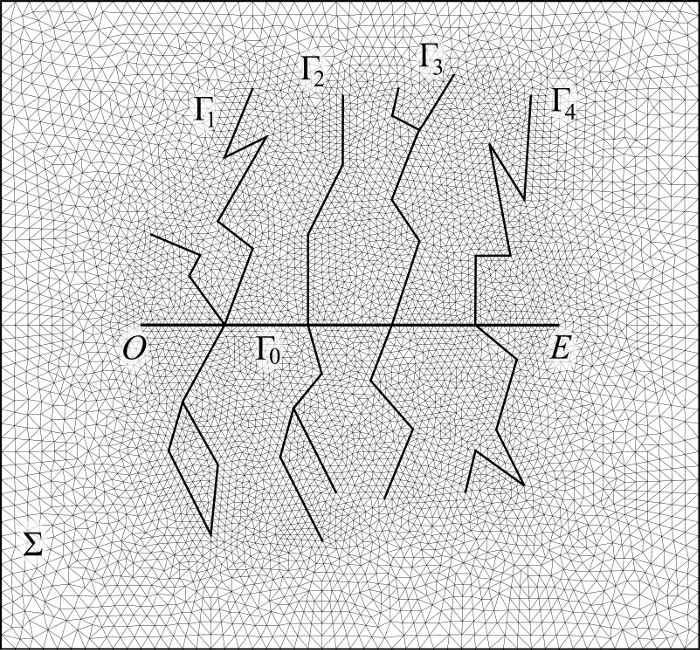}\\
  \caption{An ``arbitrary'' set of fractures and the computational mesh}\label{Case4fracs}
\end{figure}

The model allows specifying any 2D net of fractures (also with self-in\-ter\-sec\-tions) with different conductivities of its segments. In this case we observe a set of fractures shown in Figure \ref{Case4fracs}. The produced volume of fluid is compared for different rock permeabilities, reservoir heights and outer boundary conditions.  The data set enumerated by a binary 3-digits number as given in Table \ref{T4}. We use two types of outer boundary conditions: Neumann (no-flow): $\partial p/\partial n|_\Sigma=0$ and Dirichlet  (given pressure): $p|_\Sigma=0$.

\begin{figure}[t]
    \centering
  \includegraphics[width=0.8\textwidth]{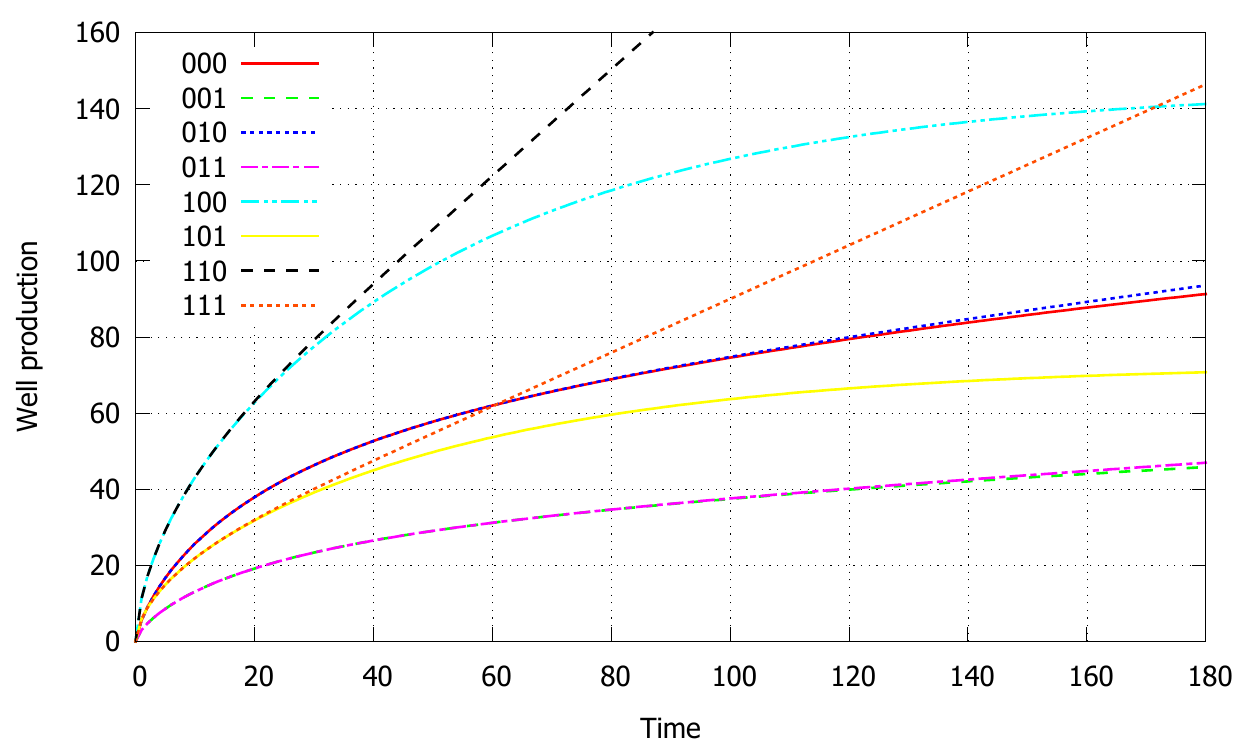}\\
  \caption{Well production ($10^3$ m$^3$) versus time (days) for an ``arbitrary'' set of fractures. Values of the instrument parameters are coded by a 3-digits binary number according to Table \ref{T4}}\label{Case4}
\end{figure}

The produced volume of fluid is shown in Figure \ref{Case4}. The difference in outer boundary conditions is clearly seen for high permeability (compare cases 100--110, 101--111). For the Neumann outer boundary condition the reservoir contains only a finite volume of fluid, which implies that for large time the produced volume is limited by the total volume in contrast with the case of the Dirichlet boundary condition where the total volume is unlimited and the produced volume tends to a linear function of time. Difference in the reservoir's heights gives the proportional increase of the productivity as is seen from the comparison of cases \#xx0 and \#xx1. The dependence on the permeability is also expectable: higher volumes of produced fluid correspond to higher permeability.

\begin{table}[t]
	\caption{Set of instrumental parameters for Case 3}
	\label{T4}
	\begin{center}
	\begin{tabular}{l p{4 cm} p{4 cm} p{4 cm}}
		\hline\hline
 		index & rock permeability $k$, mD & outer boundary conditions  & reservoir's hight $L_z$\\
		\hline
 		0 & 1  & no-flow & 20 m  \\
        1 & 10 &  given pressure & 10 m \\
		\hline
	\end{tabular}
	\end{center}
\end{table}

\begin{figure}[p]
 \begin{minipage}[t]{0.75\textwidth}
  \includegraphics[width=0.97\textwidth]{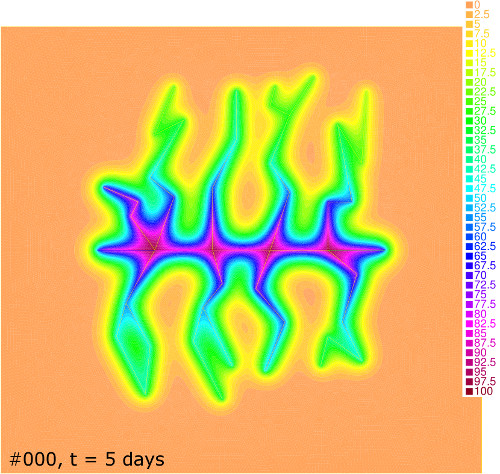}
 \end{minipage}\hfill
 \begin{minipage}[b]{0.25\textwidth}
  \includegraphics[width=\textwidth]{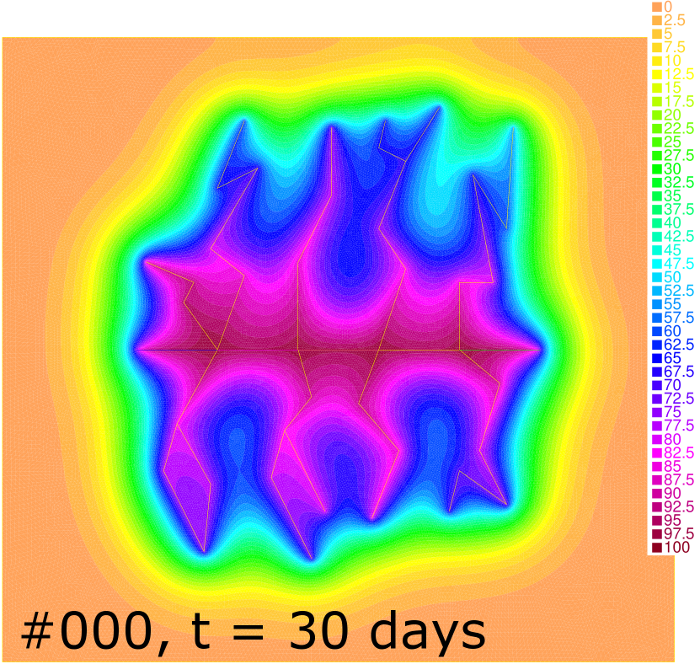}\\
  \includegraphics[width=\textwidth]{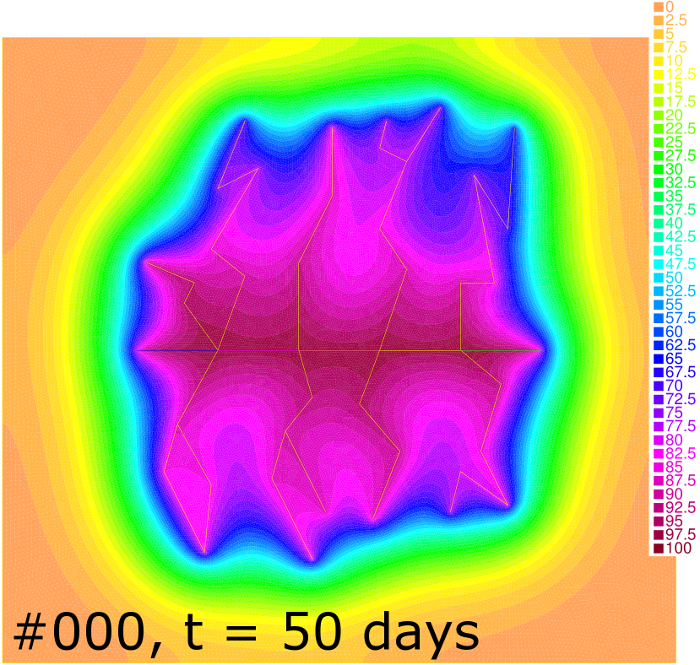}\\
  \includegraphics[width=\textwidth]{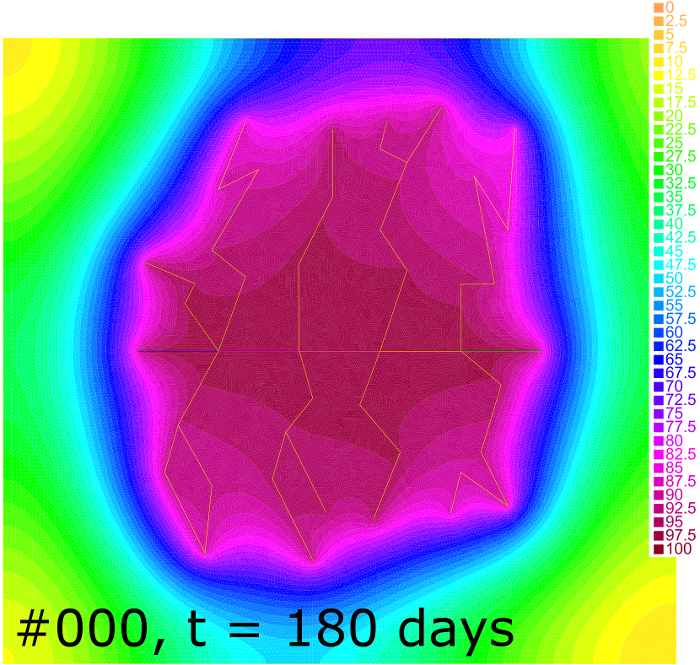}
  \end{minipage}

 \begin{minipage}[t]{0.75\textwidth}
  \includegraphics[width=0.97\textwidth]{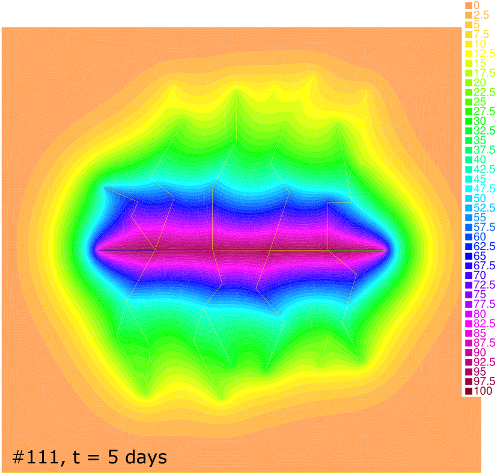}
 \end{minipage}\hfill
 \begin{minipage}[b]{0.25\textwidth}
  \includegraphics[width=\textwidth]{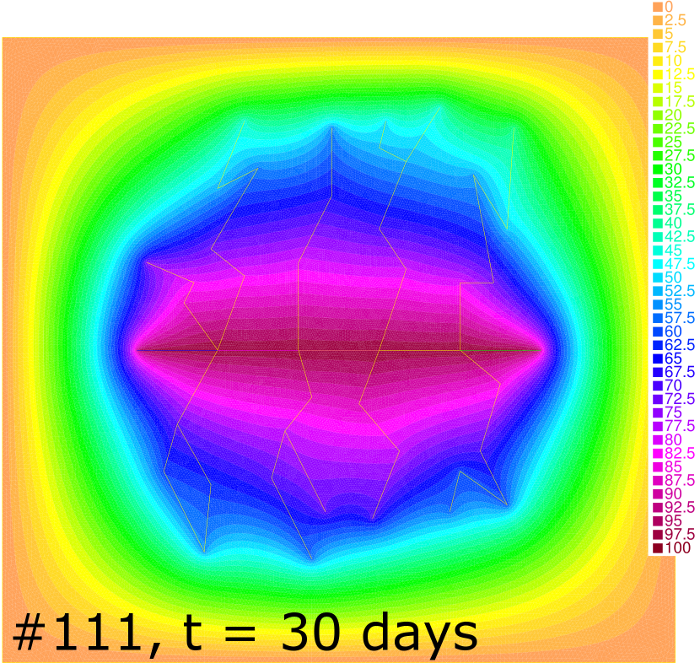}\\
  \includegraphics[width=\textwidth]{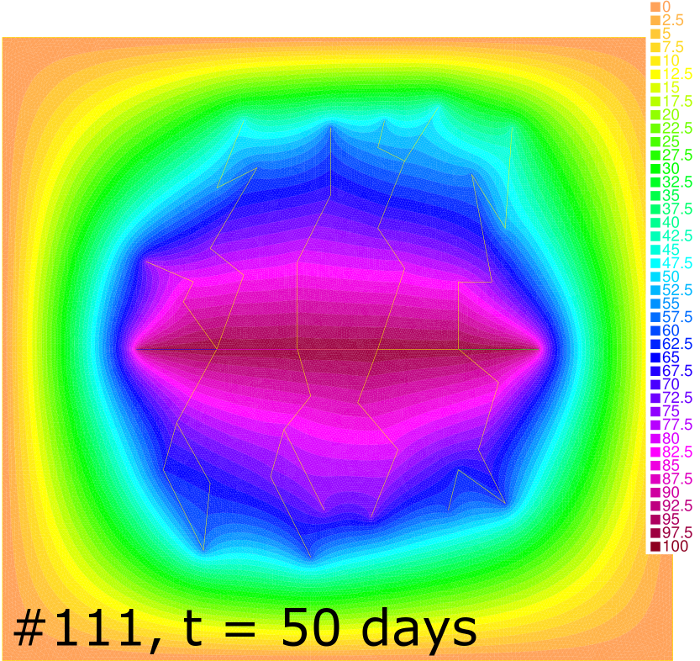}\\
  \includegraphics[width=\textwidth]{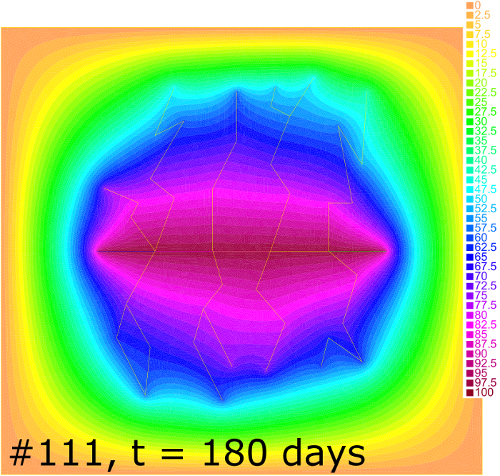}
 \end{minipage}
  \caption{Pressure distribution (atm) for cases \#000 and \#111 (Table \ref{T4}) at $t=5$, 30, 50, and 180 days.}\label{Case4Press}
\end{figure}
Pressure distribution for cases \#000 and \#111 at $t=5$, 30, 50 and 180 days is shown in Figure \ref{Case4Press}. One can see that the depression zone spreads faster in case of the higher rock permeability. The Dirichlet outer boundary condition in case \#111 allows a fluid inflow to the reservoir from the surrounding media. Together with the high permeability this brings the flow to the stationary regime starting at $t\approx 40$ h as it follows from the comparison of the bottom-right column of pictures in Figure \ref{Case4Press} and from the corresponding productivity curve in Figure \ref{Case4}. In contrast, the Neumann boundary condition in case \#000 leads to the leveling of pressure for higher $t$.

\section{Conclusion}

The mathematical model of fluid flow in a rectangular reservoir drained by a horizontal wellbore with an arbitrary network of vertical hydraulic fractures is proposed. The model allows computation of the time-dependent fluid pressure distribution within all segments of the computational domain (reservoir, fractures and wellbore) as well as integral characteristics of the flow such as the wellbore productivity. The model is applicable for reservoirs with variable in time and space physical properties (i.e. rock permeability, porosity, imperfection coefficients, etc.) and various scenarios of production (prescribed borehole pressure or flow rate as a function of time). The model does not use restrictive assumptions regarding the structure of the fluid flow and involve only a limited set of empirical parameters (wellbore and fractures imperfection coefficients). The weak formulation of the model given in the paper is suitable for the numerical solution of the model using the finite element method.

In case when all fractures extend from the bottom to the top of the reservoir, the model is reduced to a two-dimensional one by averaging along the vertical coordinate. The 2D model is implemented in a numerical code using the finite element solver FreeFEM++. Case studies for the 2D model demonstrate the dependence of the wellbore productivity on the physical and geometrical characteristics of the reservoir, of the outer boundary conditions and of the characteristics of the fracture network. This analysis suggests further applications of the model as an optimization tool for estimation of wellbore productivity under different reservoir development plans. Ability of the model to produce the fluid pressure field in the reservoir for various production scenarios allows one to compute initial data for industrial reservoir simulators. The model can also be used for the direct examination of validity of hypotheses regarding the character of fluid flow in different stages of the reservoir development, used in analytical and semi-analytical models that are cited in Introduction.

\section{Acknowledgements}
The work was partially supported by President Grant for Leading Scientific Schools of the Russian Federation (grant N.Sh.-2133.2014.1) and by RFBR (grant 16-01-00610).

\appendix
\section{Computation of coefficients $\alpha_j$ for the 2-D model}\label{App1}

\begin{figure}[t]
  \centering
  \includegraphics[width=0.5\textwidth]{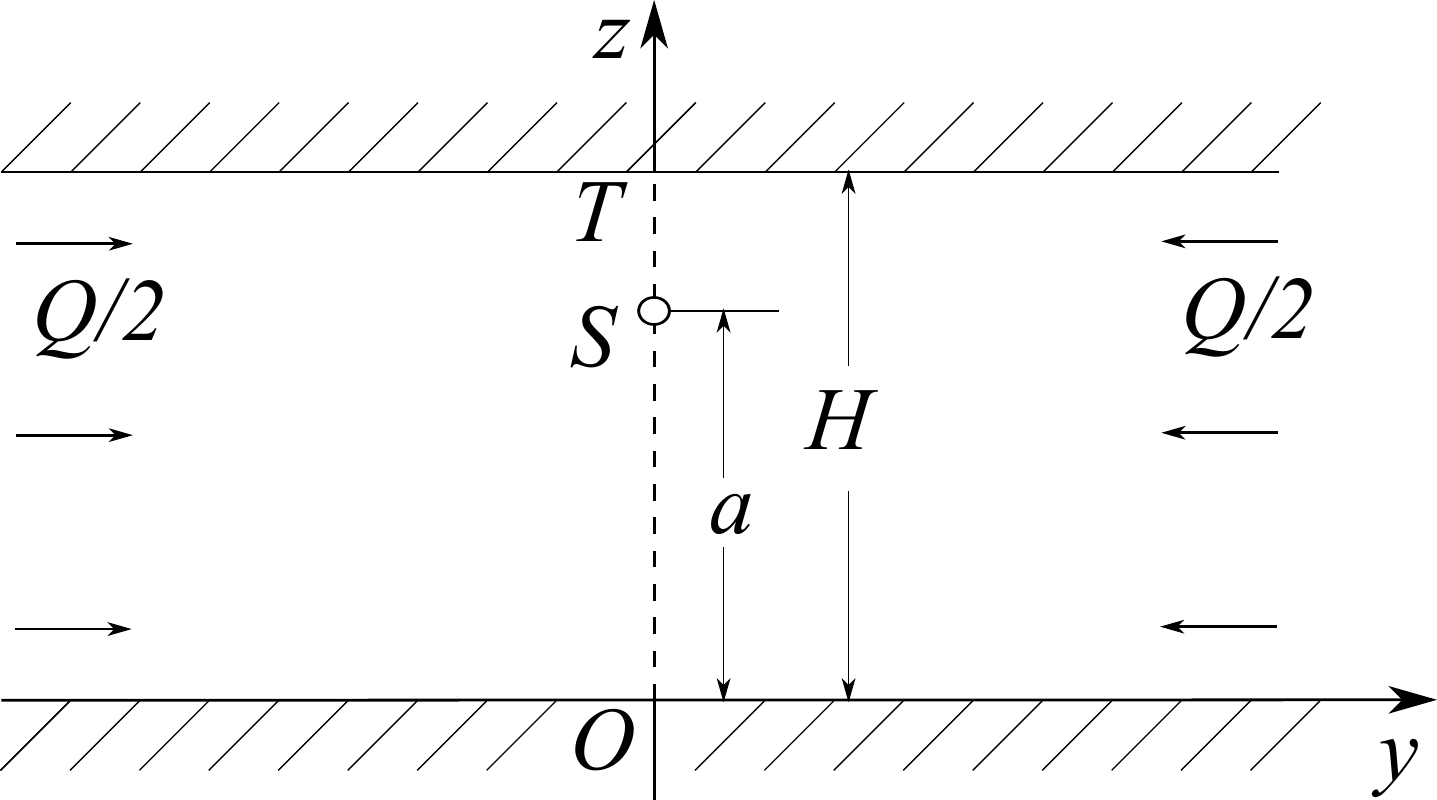}\\
  \caption{Stationary potential fluid flow in a stripe with a slit ($OT$) or a point ($S$) sink of intensity $Q$. }\label{F1}
\end{figure}

For the computation of coefficient $\alpha_j$ let us assume a problem of filtration of incompressible fluid in an infinite stripe of width $H$ between two impermeable planar walls, generated by either a point or a slit sink of a given intensity $Q$. The geometry of the problems is described in Figure \ref{F1}.

The slit sink coincide with the interval $OT$ over $Oz$-axis. The point sink $S$ is located on the hight $a$ from the bottom of the stripe.

The flow is governed by the relation $\mathbf{v}=-\sigma\nabla p$ ($\sigma = k/\mu$ for the fracture with proppant or $\sigma = d^2/(12\mu)$ for the clean fracture of width $d$) and the continuity equation $\dv \bv=0$. Thus, pressure $p$ satisfies the Laplace equation with no-flow conditions over the boundaries of the stripe. The flow rate at the infinity is given as
\[\pd{p}{z}\Bigr|_{z=0,H}=0, \quad -H\sigma\pd{p}{y}\Bigr|_{|y|\to\infty}=\frac{Q}{2}.\]
The exact solution of the problem on a slit sink in a stripe has the form
\[p_f=\frac{Q}{\sigma}\left(\frac{|y|}{2H}-\frac{\ln 2}{\pi}\right).\]
The solution of the problem with a point sink reads as follows:
\[p_s=\frac{ Q}{4\pi `\sigma}\left[\ln\left(\sinh^2\Bigl(\frac{\pi y}{2H}\Bigr)+\sin^2\Bigl(\frac{\pi(z-a)}{2H}\Bigr)\right) + \ln\left(\sinh^2\Bigl(\frac{\pi y}{2H}\Bigr)+\sin^2\Bigl(\frac{\pi(z+a)}{2H}\Bigr)\right)\right]\]
The constant arbitrariness in the definition of function $p$ is exploited to satisfy the matching condition: $\lim\limits_{|y|\to\infty}(p_f(y,z)-p_s(y,z))=0$. Let us calculate the asymptotical behaviour of function $p_s$ at $r=\sqrt{y^2+(z-a)^2}\to 0$:
\[p_s\bigr|_{r\to 0}=\frac{Q}{2\pi \sigma}\left[\ln r+\ln\left(\frac{\pi}{2H}\sin\Bigl(\frac{a\pi}{H}\Bigr)\right)\right]+\frac{Q(z-a)}{4\sigma H}\cot\Bigl(\frac{a\pi}{H}\Bigr)+O(r^2)\]
We require the following condition to be satisfied:
\[Q=\alpha\bigl(p_f|_{y=0}-{p_s}|_{r=R}\bigr)\]
accurate to the small terms of order $R/H$. After cancellation of $Q$ accurate to the small terms we obtain
\[1=\alpha\sigma\left[-\frac{\ln 2}{\pi}-\frac{\ln R}{2\pi}-\frac{1}{2\pi}\ln\left(\frac{\pi}{2H}\sin\Bigl(\frac{a\pi}{H}\Bigr)\right)\right].\]
This finally gives the formula for $\alpha$ as
\[\alpha=-\frac{2\pi \sigma}{\ln\left(\frac{2R\pi}{H}\sin\bigl(\frac{a\pi}{H}\bigr)\right)}.\]

\end{document}